\documentstyle[editedvolume,epsf,psfig]{crckapb}

\def\spose#1{\hbox to 0pt{#1\hss}}
\def\simless{\mathrel{\spose{\lower 3pt\hbox{$\mathchar"218$}}
     \raise 2.0pt\hbox{$\mathchar"13C$}}}
\def\simgreat{\mathrel{\spose{\lower 3pt\hbox{$\mathchar"218$}}
     \raise 2.0pt\hbox{$\mathchar"13E$}}}
\def\etal{{\frenchspacing\it et al.}}
\def\ie{{\frenchspacing\it i.e.}}
\def\eg{{\frenchspacing\it e.g.}}
\def\deg{^{\circ}}

\begin{opening}
\title{TOWARD HIGH-PRECISION MEASURES\protect\\
  OF LARGE-SCALE STRUCTURE}
\subtitle{}

\author{MICHAEL S. VOGELEY}
\institute{Princeton University Observatory\\
           Peyton Hall, Princeton, NJ 08544\\
	   vogeley@astro.princeton.edu}

\end{opening}

\runningtitle{High-Precision Large-Scale Structure}
\runningauthor{VOGELEY}

\begin{document}

\footnotetext[1]{Hubble Fellow}

\begin{abstract}
I review some results of estimation of the power spectrum of density
fluctuations from galaxy redshift surveys and discuss advances that
may be possible with the Sloan Digital Sky Survey.
I then examine the realities of power spectrum estimation in the presence
of Galactic extinction, photometric errors, galaxy evolution,
clustering evolution, and uncertainty about the background cosmology.
\end{abstract}

\vskip-0.5cm

\section{INTRODUCTION}

\vskip-7.30truecm
\hskip6.1truecm{$^1$}
\vskip7.0truecm

\vskip-12.00truecm
{$\>$}\\
{\it 
To appear in {``Ringberg Workshop on Large-Scale Structure''},\\
ed. D. Hamilton (Kluwer, Amsterdam)
}
\vskip11.00truecm

The advent of deep redshift surveys of $10^4-10^6$ galaxies, such as
the Las Campanas Redshift Survey (LCRS; Shectman \etal~1996) and the
AAT 2df survey (Colless 1998), and multiband photometric and
spectroscopic surveys such as the Sloan Digital Sky Survey (SDSS; Gunn
\& Weinberg 1995; Szalay 1998), marks the beginning of a new era of
investigations in large-scale structure.  Rather than treat galaxies
as indistinguishable tracers of mass in a static distribution, we will
study the detailed dependence of clustering on galaxy type and cosmic
epoch.  In fact, as I will discuss, we must study the species and
redshift dependence of clustering to precisely differentiate among
cosmological models. In somewhat shallower wide-angle surveys (Huchra
\etal~1983; Giovanelli \& Haynes 1984; Geller \& Huchra 1989; da
Costa \etal~1994; Fisher \etal~1995) it has been possible to begin
study of these effects; at redshift $z=0.1$ and larger it becomes
critical to do so.  In addition to precisely characterizing galaxy
clustering at the present epoch, we will use the anisotropy of
clustering, as well as the evolution of this clustering, to constrain
cosmological parameters.

A key issue that I examine is the limiting precision with which we can
hope to measure galaxy clustering at the present epoch.  Although much
has been made of the power of large galaxy surveys to measure
cosmological parameters, most of this discussion has focussed on
finding an analysis method that gives ``optimal'' uncertainties on
model parameters, where the uncertainties are due to cosmic variance
and shot noise (Vogeley \& Szalay 1996; Tegmark \etal~1998).
Much less has been said about the uncertainties caused by
random and systematic errors in photometry or Galactic extinction
corrections, and systematic uncertainties that arise from issues of
cosmology, and evolution of galaxies and their clustering.  In this
paper I discuss the impact of these effects on analysis of galaxy
redshift samples from the SDSS.

\section{MEASUREMENTS OF THE POWER SPECTRUM FROM GALAXY SURVEYS}

Rather than give a review of the myriad measures of galaxy clustering
that we hope to apply to these new surveys, here I focus on one
statistic and examine how possible random and systematic
uncertainties will affect the accuracy of this
measurement.  Not surprisingly, I consider the power spectrum of
galaxy density fluctuations, because lowest-order measures are where
we first gain precision, and because uncertainties in higher-order
measures often scale from the uncertainties in the variance.

\begin{figure}
\epsfxsize=12.5cm
\epsfbox{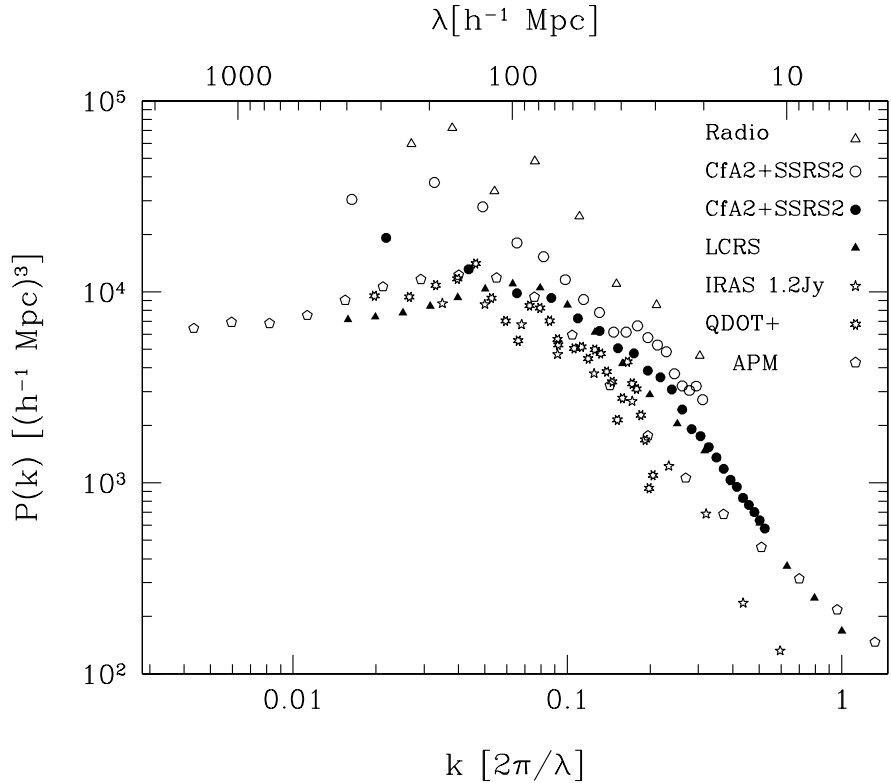}
\vskip-0.5cm
\caption{
Estimates of the redshift-space power spectrum from a variety of redshift
surveys, and an estimate of the real-space power spectrum inferred from 
angular correlations (APM).
}
\end{figure}

\subsection{THE DATA}

Figure 1 plots the power spectrum $P(k)$ in units of $h^{-3}$ Mpc$^3$
(a power spectral density) as estimated from a number of radio
(Peacock \& Nicholson 1991), infrared (Fisher \etal~1993; Tadros \&
Efstathiou 1995), and optically-selected (da Costa \etal~1994; Lin
\etal~1995) galaxy redshift survey samples, and the real-space power
spectrum inferred from angular clustering of a photometric sample
(Baugh \& Efstathiou 1993).  To avoid hopelessly confusing the plot,
no error bars are shown (unfortunately, the various authors differ
somewhat in their choice of binning and methods of computing these
uncertainties, so comparison of these errors would not be as
informative as one might hope). The shapes of most of these spectra
roughly agree over the decade $10-100h^{-1}$ Mpc.  However, there is
large variation of their amplitudes. We usually attribute this variation to
differences in the ``bias parameter,'' which affects the amplitudes as
$P_1(k)/P_2(k)=(b_1/b_2)^2$ if the bias is scale-independent.  This
strong dependence of clustering amplitude on method of galaxy
selection implies that detailed knowledge of the selection criteria
and construction of sub-samples for analysis that are homogeneous in
this selection are prerequisites for precision measurement of the
power spectrum.

To clarify comparison of the shapes of the power spectra, Figure 2
plots these same data, with the amplitudes of the curves shifted so
that all match $P(k)$ of the LCRS sample near wavenumber $k=0.1$.  No
shift was required to obtain agreement between the LCRS and the
CfA2+SSRS2-100 volume-limited sample, which is not surprising since
they both include optically-selected galaxies that are roughly $M^*$
and brighter.  In general, one sees excellent agreement between optical
redshift samples for wavenumbers $k>0.1$.  There is some disagreement
between infrared and optical samples over this same range of wavenumber,
perhaps because the selected galaxies sample different physical
environments.  One should be careful not to over-interpret this
comparison of spectral shapes.  Where large shifts are necessary to
match the amplitudes, this shifting procedure is questionable, because
the effects of redshift distortions (both the linear boost described
by Kaiser 1987 and the washing out of small-scale power due to
redshift fingers of clusters) and non-linear evolution both depend on
the ratio of galaxy-clustering to mass-clustering amplitude.  On large
scales, at wavenumber $k<0.1$, we find significant departures in the shapes
of the spectra, with a scatter of roughly a factor of three among the
samples.  This scatter is consistent with the uncertainty due to
cosmic variance (small sample volume relative to the wavelength scales
of interest).

\begin{figure}
\epsfxsize=12.5cm 
\epsfbox{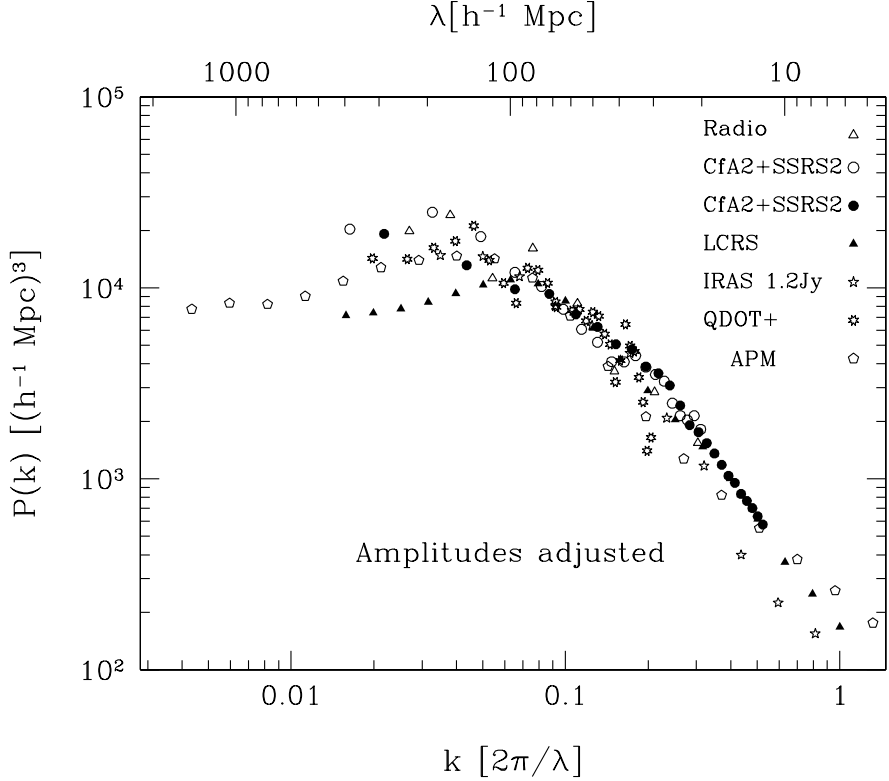}
\vskip-0.5cm
\caption{Power spectrum estimates from Figure 1 arbitrarily scaled to match at
$k=0.1$ to allow comparison of power spectrum shapes.
CfA2+SSRS2-101 and LCRS were not adjusted.
}
\end{figure}

\subsection{MODEL FITS}

The shape of these spectra may be well-fitted by a linear Cold Dark
Matter (CDM) power spectrum (Bardeen \etal~1986) with shape parameter
$\Gamma=0.25$ (where $\Gamma=\Omega h$ in the simplest
models, with $h=H_0/100$). 
Here we use the linear
CDM spectrum merely as a fitting function. It is interesting to note
that, for values of $\Gamma=\Omega h \sim 0.25$, this linear curve
does a reasonable job in fitting the redshift-space power spectrum of
a fully non-linear evolved CDM model because the effects of non-linear
evolution and redshift distortion nearly cancel out.  

\begin{figure}
\epsfxsize=12.5cm 
\epsfbox{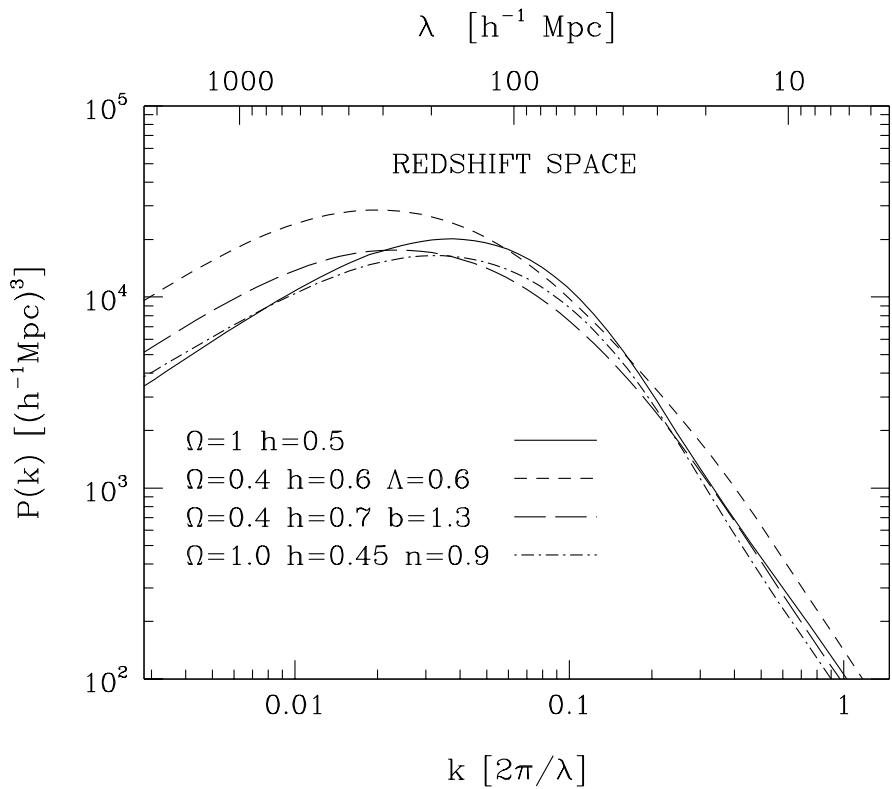}
\vskip-0.5cm
\caption{Non-linear redshift-space power spectra for a variety of
COBE-normalized CDM models.}
\end{figure}

Does this fit to the spectral shape imply that our work is done?
Hardly. This rough shape might be matched by tweaking the parameters
of several competing models (\eg, by adding some cosmic density in
neutrinos and/or by varying the slope of the primordial power
spectrum).  We require higher resolution to detect features in the
power spectrum that might be caused by acoustic oscillations near
recombination (Eisenstein \& Hu 1998) or other physical effects.  Higher
resolution in the Fourier domain requires a larger survey volume (see
next section), which is also necessary to probe large wavelength
scales. Measurements of the power spectrum to accuracies of a few
percent on scales $\lambda > 100 h^{-1}$Mpc would allow detailed
comparison of clustering of galaxies at the present epoch with
clustering of the mass at redshift $z=10^3$, as revealed by the
anisotropy of the Cosmic Microwave Background (CMB).

\begin{figure}
\epsfxsize=12.5cm 
\epsfbox{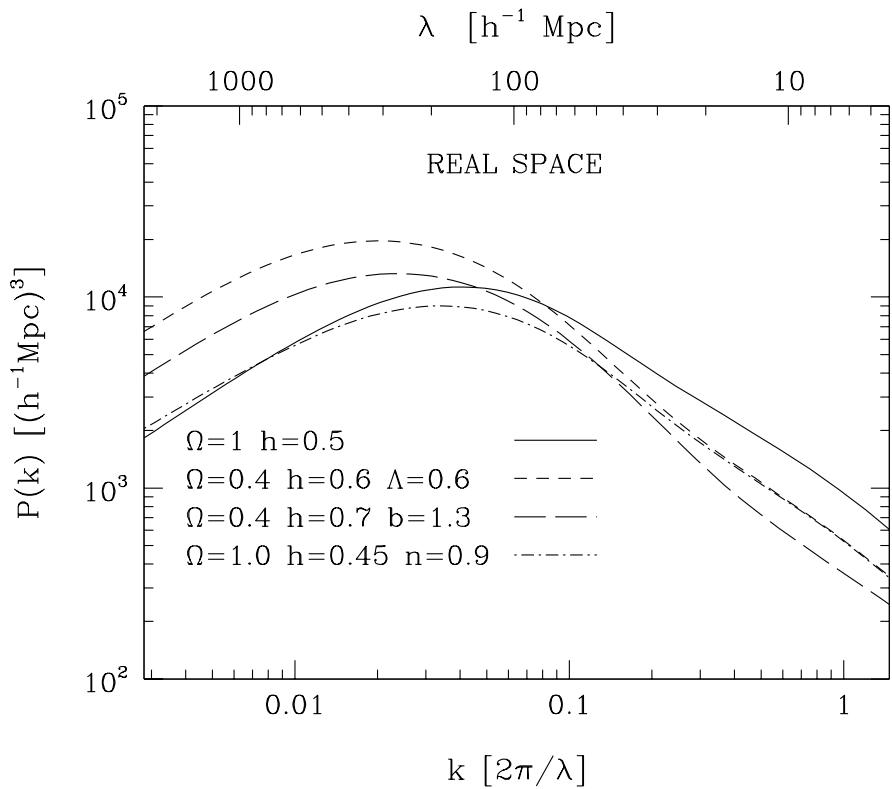}
\vskip-0.5cm
\caption{Real-space power spectra for the same models plotted in Figure 3.
}
\end{figure}

Figure 3 illustrates how CDM models with different parameters yield
redshift-space power spectra that are similar.  To compute these
spectra, I use the semi-analytic formalism described by Peacock (1997)
to approximate the effects of non-linear gravitational evolution and
redshift distortions on the linear CDM spectrum. All of these models
have a mass spectrum that is normalized to match the COBE DMR
measurement of the CMB anisotropy (Wright \etal~1996). Even without
exploring admixtures of CDM and neutrinos (\eg, Primack \etal~1995),
we see that several choices of parameters might fit the equally well
(or equally poorly, depending on one's taste). Not examined here is
how the small-scale power spectrum depends on the details of galaxy
formation relative to the mass distribution.

\subsection{REAL VS. REDSHIFT SPACE}

One of the problems of working in redshift space is that a large
amount of small-scale power creates a large small-scale velocity
dispersion. In redshift space, the resulting ``fingers of God'' wash
out much of this same small-scale power. Similarly, for fixed
normalization and Hubble constant, CDM models with large $\Omega$
experience a significant boost of their redshift-space power spectrum
amplitudes (1.87 if $\Omega=1$ and $b=1$), which makes their
large-scale spectra look more like those of low $\Omega$ models.

Redshift-space power spectrum measurements with accuracy of a few
percent could differentiate between the model spectra in Figure 3
but, clearly, it would be easier to break the model degeneracy that we
see in redshift-space power spectra if we could access the underlying
real-space (\ie, configuration space) power spectrum of the galaxy
distribution.  Figure 4 shows the power spectra of the same models
from Figure 3, now plotted in real space.

The lesson here is not that we should use the angular correlation
function, and give up measuring redshifts, but rather that we should
take advantage of the anisotropy of clustering in redshift space to
allow measurement of both the power spectrum and $\Omega$.  Andrew
Hamilton's contribution to this volume examines this issue
in great detail.

\section{MINIMAL POWER SPECTRUM UNCERTAINTIES FOR THE SDSS}

Power spectrum measurements with accuracy of a few percent on scales
from one to a few hundred Mpc would allow us to make great strides in
constraining cosmological models.  Is this feasible? Here I begin to
examine the accuracy with which we dare hope to measure $P(k)$ in the
foreseeable future. 

\subsection{APPROXIMATE UNCERTAINTY PREDICTIONS}

\begin{figure}
\epsfxsize=12.5cm 
\epsfbox{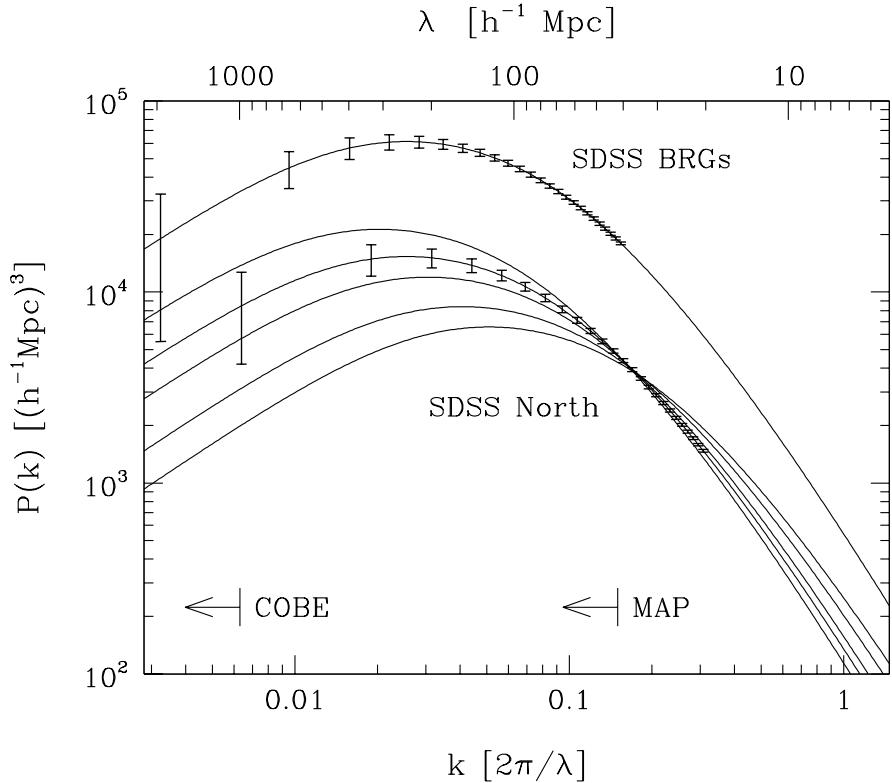}
\vskip-0.5cm
\caption{
Predicted uncertainties in the power spectrum estimated from a
volume-limited ($R_{max}=500 h^{-1}$ Mpc) sample of SDSS North and for
the Bright Red Galaxy sample (upper set of error bars). These errors
assume that the true power spectrum is that of an $\Omega h=0.25$ CDM
model and that the BRGs are more clustered than normal galaxies.
Plotted for comparison to the SDSS North errors are CDM power spectra
(normalized to $\sigma_8=1$) for a range of $\Omega h$ from $0.2$
(uppermost curve) to $0.5$ (lowest curve) and indications of the range
of comoving scales probed by the COBE and MAP CMB anisotropy
experiments.  
}
\end{figure}

For fixed survey strategy, the survey volume and the number density of
galaxies in the redshift sample set a lower bound on the uncertainty
in the estimated power spectrum.  In the limit of a perfectly
spherical volume-limited sample, the uncertainty in the estimated
power per mode is roughly
\begin{equation}
{\delta \tilde{P}(k)\over \tilde{P}(k)}  \approx \sqrt{2{V_c \over V_k}}
\left[1+ {S(k)\over P(k)} \right]
\end{equation}
(Feldman, Kaiser, \& Peacock 1994).  $P(k)$ and $\tilde{P}(k)$ are the
true and estimated power spectra, respectively.  $S(k) = 1/\bar{n}$ is
the shot noise power for mean galaxy density $\bar{n}$.  $V_c =
{(2\pi)^3/ V_S}$ is the coherence volume in the Fourier domain for a
survey with volume $V_S$.  It is assumed that we average the
power estimates over a shell in Fourier space with volume $V_k \approx
4\pi k^2 \Delta k$. That is, we average the power over all angles, and
over bins with width $\Delta k > 2\pi/R$, where $R$ is the survey
depth.  For a larger volume, the coherence volume in Fourier space
decreases, thus we obtain a larger number of independent probes of the
power spectrum.  This convenient approximation breaks down when the
survey volume is highly anisotropic; in this case the coherence volume
in Fourier space (the ``window function'') becomes anisotropic and the
range of true wavenumber that is probed depends on the direction of
$\bf{k}$.
Figure 8 of Alex Szalay's contribution to this volume
shows window function shapes and relative sizes for a variety of
redshift surveys, including the SDSS.

A volume-limited sub-sample of the SDSS redshift survey in the North
Galactic Cap region (this covers solid angle of $\pi$ steradians) that
has comoving coordinate depth of $R=500h^{-1}$ Mpc will include all
galaxies with absolute magnitude roughly $0.4^m$ brighter than $M^*$,
with mean number density $\bar{n}\sim 2\times 10^{-3} h^3$Mpc$^{-3}$.
Hereafter I will use this as a fiducial sample for computing
uncertainties in the estimated power spectrum.  This sample is
conservative, in the sense that the SDSS will allow selection of much
deeper samples, which include more volume, and because a
magnitude-limited sample to the same depth would have larger average
number density, hence smaller shot noise.  

Figure 5 plots $1\sigma$ uncertainties for power estimates binned by
$\Delta k=2\pi/ (500 h^{-1}$Mpc) and assuming that the true power
spectrum is described by a $\Gamma=0.25$ linear CDM spectrum.  Also
shown are a family of CDM spectra with different shape parameter
$\Gamma$. The uncertainty in each bin of width $\Delta k$ at a
wavelength scale of $\lambda=100h^{-1}$Mpc will be
$\sigma(\hat{P})/\hat{P}\approx 0.06$.
Such a measurement would easily differentiate between
various CDM models, as the family of CDM curves indicates.  Comparison
with results from the LCRS further llustrates the power of such a
large survey: if a ``bump'' at $\lambda\sim 128h^{-1}$Mpc were found
in the SDSS sample with $\delta \hat{P}/\hat{P}_{smooth}=0.76$ in a single bin
(as seen by Landy \etal~1996), this would be a $12\sigma$ event.

\subsection{SAMPLES WITH LARGER VOLUME?}

Could we do better than the SDSS with, for example, an all-sky survey
to similar depth? This suggestion is not out of the question, if we
combine all of the redshift surveys that will exist in a few years'
time.  The SDSS NGP survey will cover one quarter of the sky; a
similar full-sky survey would have power spectrum uncertainties that
are at best a factor of two smaller.
More realistic at optical wavelengths would be a survey
over Galactic latitude $|b|>30$, which covers half the sky, yielding
errors that are at best $40\%$ smaller than the SDSS. In fact, the
SDSS itself will cover an effective area larger than $\pi$ steradians,
by including three slices in the southern Galactic hemisphere, each
$3^{\circ}\times \sim 100^{\circ}$ in area. If the SDSS is combined
with the AAT 2df redshift survey and the LCRS, then most of the high-latitude
sky will be sampled.  Galaxy selection in the infrared would allow a
true full-sky survey, but no IR photometric survey exists to depth
comparable to the SDSS.  The IRAS PSC-Z survey (Saunders \etal~1994)
comes closest to this goal, but has much larger shot noise
(smaller galaxy density) and apparently more severe sample evolution.

One-tenth of the million galaxies targeted for spectroscopy by the
SDSS will be a sample of Bright Red Galaxies (BRGs) that are
volume-limited to redshift $z=0.45$. The volume limit will be enforced
by selecting the galaxies by their absolute magnitudes, estimated
using the photometric redshift technique (Connolly \etal~ 1995).  This
sample will cover a volume eight times larger than the fiducial
SDSS-500 sample described above, albeit with larger shot noise.  The
upper curve and error bars in Figure 5 show the expected uncertainties
from cosmic variance and shot noise in the SDSS BRG sample, using
equation 1 and assuming that the BRGs are biased by a factor of 2
(factor of 4 in the power spectrum) relative to the normal galaxy
sample.  The brightest and reddest galaxies appear to evolve least of
all, so this sample will provide our best chance to measure the
evolution of clustering (see section 5.3 below).

\section{MEASUREMENT ERRORS AND EXTERNAL SYSTEMATICS}

In the previous section I describe ``ideal'' uncertainties that are
lower bounds on the total error budget. I now examine whether these
expectations are realistic. What other effects limit our knowledge of
clustering at the present epoch?  I use the SDSS as a test case, but
the following analysis is also relevant for other surveys to similar
depth.

There are several possible effects about which we do not need to worry
for the SDSS.  Because the galaxies in the spectroscopic sample will
be five magnitudes brighter than the point source detection limit,
star-galaxy separation will not be a problem.  Galaxies with very low
central surface brightness will be excluded from the redshift sample,
because we would not be able to get redshifts of these objects, but we
will understand this selection bias quite well. For example, our
magnitude and surface brightness cuts will be ``fuzzy,'' such that we
observe some fraction of objects beyond the nominal cuts.  The
redshifts that we do measure will have $20-30~$km~s$^{-1}$ accuracy, so
this project is overkill for simply measuring redshifts and we expect
very few failures. The use of a fiber-fed spectrograph with plugplates
to hold the fibers in the focal plane imposes a 55
arcsecond minimum separation for pairs of objects, except where a
target lies in the overlap between two or more plugplate
fields. However, during the Fall observing season, the larger ratio of
spectroscopic time to photometric time (because we will image a
smaller area), will allow a larger covering factor for spectroscopy,
thus we will obtain spectra for at least all pairs of objects and
potentially could observe all n-tuples up to $n=7$. From these
southern data, we will study the effects of excluding close pairs from
the northern spectroscopy (\eg, we can study the very small-scale
velocity dispersion of galaxies from the southern spectroscopy).

\subsection{EFFECTS OF EXTINCTION AND PHOTOMETRIC ERRORS}

Despite our best efforts at correcting for Galactic extinction and
calibrating the photometry, errors in these will add to the minimal
uncertainties illustrated in Figure 5.  In collaboration with Andy
Connolly at Johns Hopkins, I have investigated the impact of such
errors on power spectrum estimation from samples of the SDSS (Vogeley
\& Connolly 1998). The remainder of this section is an overview of
some of our results.

Errors in either photometry or the extinction corrections affect the
apparent number density of galaxies in similar fashion.  Here we
consider only the effects of extinction or photometric error on total
magnitude selection.  Although the SDSS spectroscopic samples will
also be chosen on the basis of a central surface brightness cut,
simulations of the SDSS indicate that surface brightness dimming from
extinction will affect an insignificant number of galaxies.

\begin{figure}
\epsfxsize=12.5cm
\epsfbox{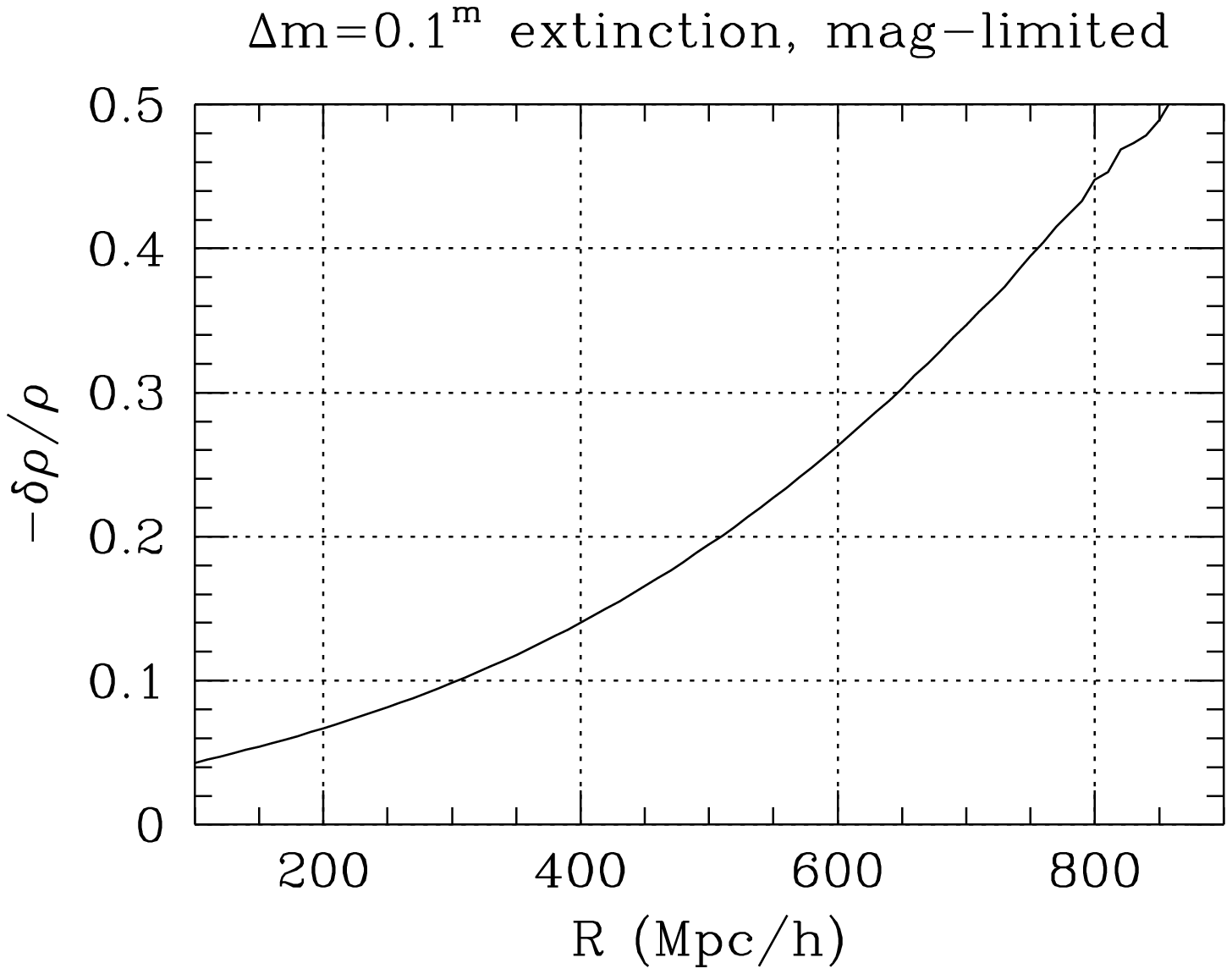}
\vskip-1.9in
\epsfxsize=12.5cm
\epsfbox{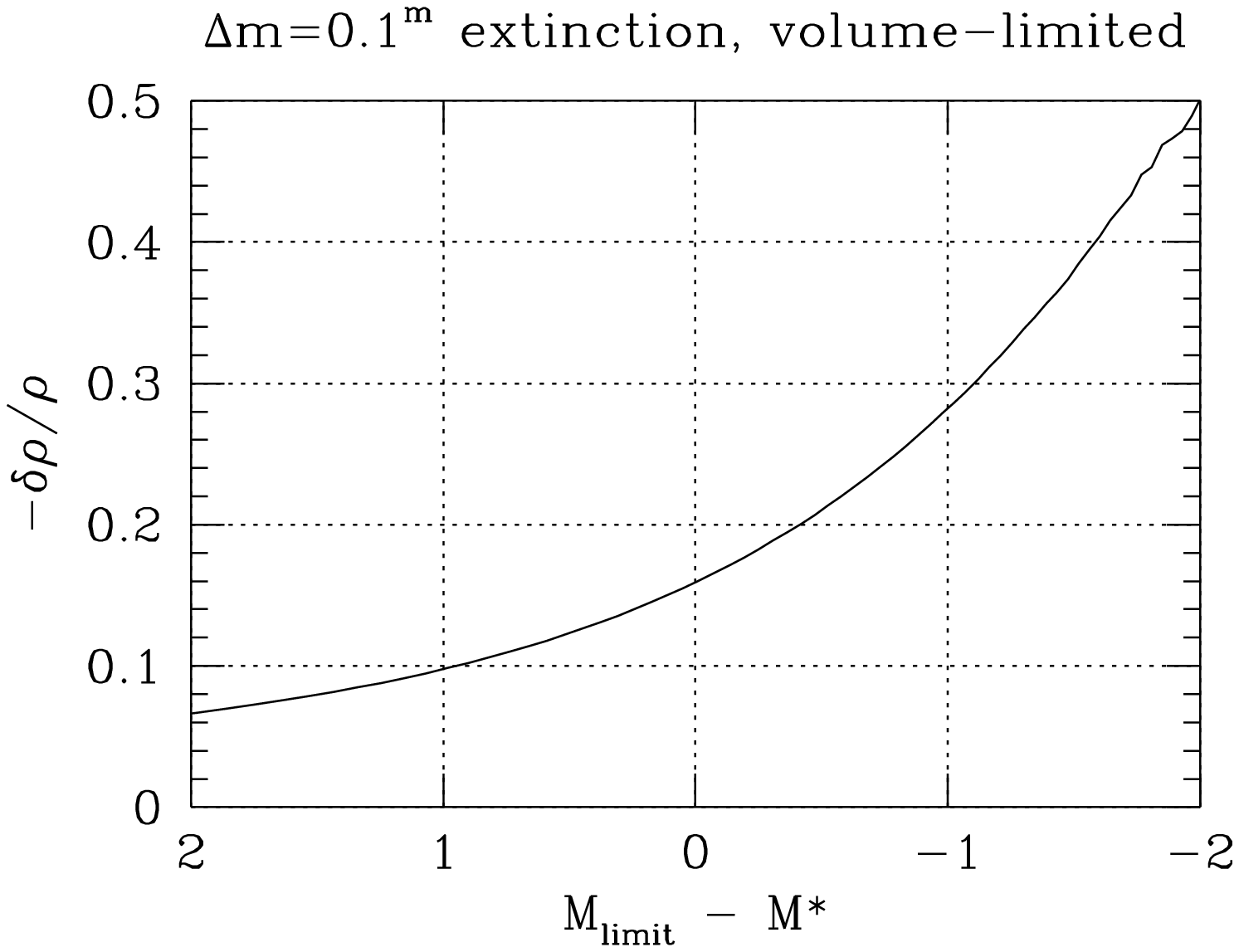}
\vskip-1.7in
\caption{Effects on the observed galaxy density that would result from
$0.1^m$ of extinction in a magnitude-limited (upper panel) or
volume-limited (lower panel) redshift sample of the SDSS.
}
\end{figure}

In a volume-limited sample that should include all galaxies brighter
than absolute magnitude $M_{lim}$, a magnitude error of $\Delta m$
over a patch on the sky causes the apparent galaxy density in that
direction to differ from the true galaxy density by
\begin{eqnarray}
{n_{obs}\over n_{true}} & = & {\Phi[< M -\Delta m] \over \Phi[< M]} \\
 & \approx & 1 - {\phi(M)\Delta m \over \Phi[<M]} \nonumber \\
 & \sim    & \exp\left [-\Delta m 10^{0.4(M*-M_{lim})}\right ], \nonumber
\end{eqnarray}
where $\phi(M)$ and $\Phi(<M)$ are the differential and integrated
luminosity functions of the galaxies.  The approximations are valid
for a Schechter luminosity function when $M_{lim}$ is close to or
brighter than $M^*$.  This form clearly shows how the strength of the
effect depends on the slope of the luminosity function near the
magnitude limit; deeper sub-samples will be more severely affected by
extinction or photometric uncertainty than shallow samples because the
count slope is steeper.  For a magnitude-limited sample with limit
$m_{lim}$, the absolute magnitude limit varies with distance
$M_{lim}(r)$ and the apparent density differs from the true galaxy
density by
\begin{equation}
 {n_{obs}\over n_{true}} = { \Phi[<M(r)-\Delta m] \over \Phi[<M(r)] }. 
\end{equation}
Figure 6 shows the effect of a $\Delta m=0.1^m$ error on the apparent
density of volume-limited and magnitude-limited samples of the SDSS.
This modulation of the apparent density grows with distance in a
magnitude-limited sample and is, on average, less severe than in an
volume-limited sample that is cut off at the same depth.  In the
following analysis I consider volume-limited samples because (1) the
effects are more severe for these, thus the analysis is conservative
in the sense of giving worst-case errors and (2) we won't believe the
results of magnitude-limited power spectrum analyses until we
understand how the clustering amplitude differs among galaxy species
(see section 2.1 above).

\subsection{THE POWER SPECTRUM OF GALACTIC EXTINCTION}

\begin{figure}
\epsfxsize=12.5cm 
\epsfbox{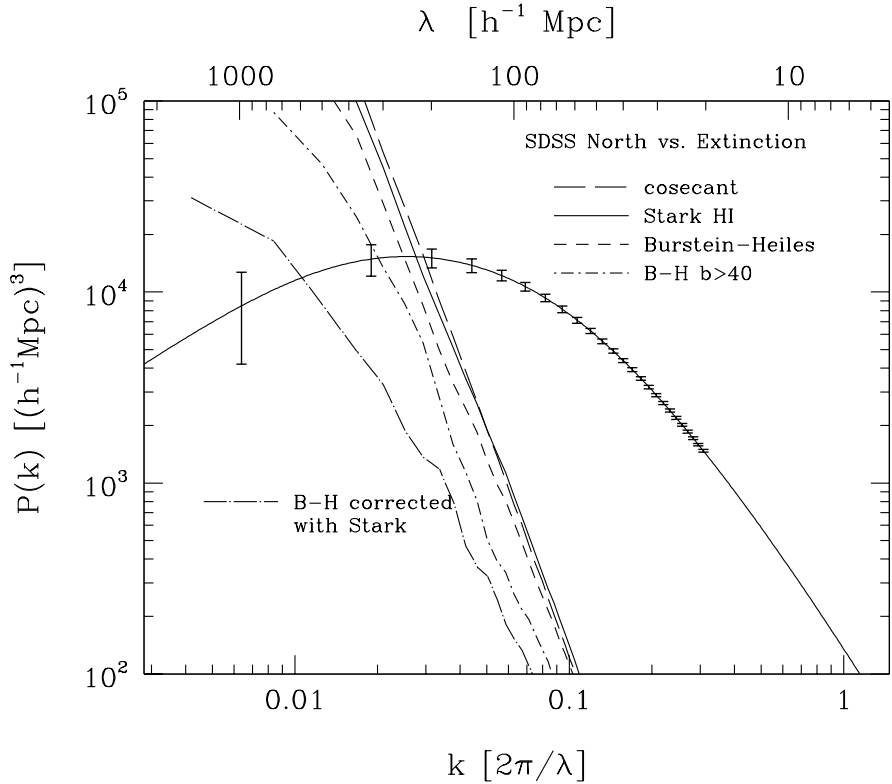}
\vskip-0.5cm
\caption{
False observed clustering power 
that would be caused by Galactic extinction, for different
assumed extinction maps. For comparison is an $\Omega h=0.25$ CDM power
spectrum and the expected uncertainties for a volume-limited sample of
SDSS North. If we make no correction for extinction, then the false clustering
power exceeds the true power for $\lambda\simgreat 300 h^{-1}$ Mpc.
Cutting back the edges of the survey to $b>40^{\circ}$ has only a small effect.
If we correct the B\&H map using a crudely-calibrated map derived from
the Stark $et~al.$ HI maps, the residual power exceeds the true power for
scales $\lambda\simgreat 700h^{-1}$ Mpc.
}
\end{figure}

Because Galactic extinction modulates the apparent density of galaxies,
it could cause erroneous clustering power to appear in our data. To
leading order, the net effect is to add to the apparent clustering,
$P_{obs}(k)\approx P_{true}(k)+P_{extinction}(k)$.  If we make no
correction at all for Galactic extinction, we expect to observe extra
clustering power in the SDSS-500 sample, as shown in Figure 7.  Here
we predict the effect of extinction over the SDSS North region using a
cosecant law, the Burstein \& Heiles (1982) map, and the Stark 
\etal~(1992) HI map.  These maps predict similar clustering power; it is
the large-scale cosecant-like variation in extinction that causes the
trouble.  Restricting the analysis to somewhat higher Galactic
latitude ($|b|>40^{\circ}$) only slightly ameliorates the problem.
Failure to notice that we live in a Galaxy that is laced with dust
would cause one to infer that the power spectrum of the universe
suddenly rises in power-law fashion beyond $200-300h^{-1}$ Mpc.

Fortunately, we are not oblivious to the effects of dust and we plan
to make some correction for this effect. The SDSS will select
spectroscopic targets after applying an {\it a priori} correction for
extinction to the apparent magnitudes, thus constructing a sample that
is uniform in an extragalactic sense.  As of this date we plan to use
the extinction maps that have been constructed from the DIRBE and IRAS
satellite data by Schlegel, Finkbeiner, \& Davis (1998).  If we or
others later construct a better extinction map, we will take the
residuals into account when we analyze the data. The erroneous
clustering power that we really need to worry about is caused by
unknown residuals between our best extinction map and the true
Galactic extinction.

The lower dot-dashed line in Figure 7 shows what happens if the true
extinction is described by the Burstein-Heiles map and we use a
crudely-calibrated (here we applied our own {\it ad hoc}
dust-to-gas relation) version of the Stark \etal~HI map to estimate
and correct for the extinction. Even with this roughly calibrated map,
the extinction power is below the true power on scales smaller than
$700h^{-1}$Mpc.  A well-calibrated map would allow us to accurately
probe the entire range of accessible scales.

\subsection{THE POWER SPECTRUM OF PHOTOMETRIC ERRORS}

\begin{figure}
\epsfxsize=12.5cm 
\epsfbox{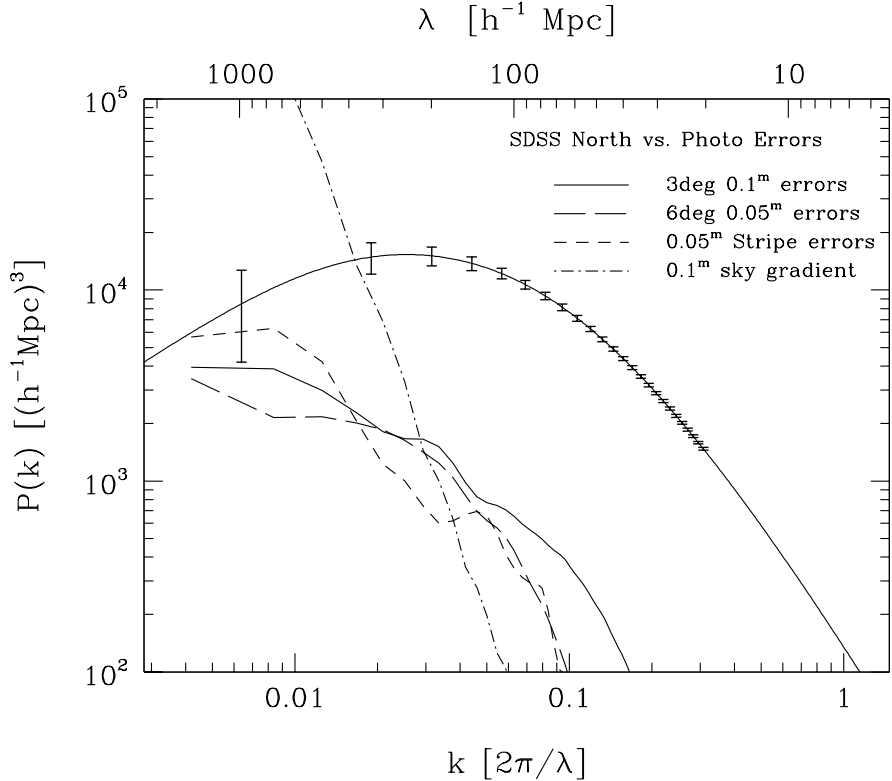}
\vskip-0.5cm
\caption{
False clustering power due to fluctuations of the photometric zeropoint
(or errors in the extinction map)
on $3\deg$ and $6\deg$ scales, or between stripes of the SDSS. 
Also shown is the effect of a $0.1^m$ gradient across the sky.
}
\end{figure}

Similar to the effects of extinction, errors in photometry that are
correlated over a patch of sky will modulate the apparent density of
galaxies. Random (uncorrelated between galaxies) errors will contribute to the
notorious Malmquist bias; here we consider only correlated photometric
errors, such as those that might arise from variations in the zero point.

Figure 8 shows the extra clustering power in the SDSS-500 sample that
we would observe if the photometric zero-point varies by $0.1^m$ over
patches that are $3^{\circ}$ in diameter (the field of view of the
SDSS camera), or by $0.05^m$ over patches that are $6^{\circ}$ in
diameter (the size of a Schmidt plate). These curves should match on
scales much larger than the patch size, but vary somewhat because we
compute these curves using a Monte Carlo realization of the errors.
Even such gross errors in photometry produce erroneous clustering that
is less than $10\%$ of the true power for $\lambda < 10^3 h^{-1}$Mpc.

The relatively small amplitude of power caused by even $0.1^m$
photometry errors over the field of view of the SDSS camera is very
good news for the extinction problem. This result implies that we can
cure the extinction problem if we can use the SDSS data themselves to
construct an extinction map that has random field-field errors that
are smaller than $0.1^m$ (but, of course, without any large-scale
systematic errors).  Variations in faint galaxy counts and colors and
colors of hot Galactic halo subdwarfs offer several means for
constructing such a map, which would provide an independent method for
computing extinction corrections.

The lesson of the extinction errors is that large-scale gradients,
rather than small-scale random photometry errors, are the real worry.
Figure 8 also shows the extra clustering power that arises if there is
a zero-point gradient across the sky (pole to pole) of $0.1^m$ or if
the zero-point randomly varies by $0.05^m$ between the ``stripes'' of
the SDSS.  In other words, the latter illustrates what happens if the
photometry is consistent within each $\sim 3^{\circ}\times
120^{\circ}$ scan region, but the calibration zero-point varies
between disjoint regions.  Even such a severe error (which we would no
doubt notice by comparing the overlap regions of the scans) would
produce extra power that is below $10\%$ of the true power on scales
$<10^3h^{-1}$Mpc. To be certain, we would not be happy to have a
$10\%$ effect in our measurements. The point is that this is clearly
an upper bound on the magnitude of such an effect.

It is important to note that, just as examination of the anisotropy of
clustering will reveal the effects of redshift distortion on the power
spectrum, comparison of clustering in the angular and radial
directions will provide a test of false clustering due to extinction
or photometric error.  Cross-correlation of the angular clustering of
samples that are selected to lie at different distance would clearly
reveal the signature of correlated magnitude errors.
\newpage

\section{EVOLUTION: THE UNIVERSE, GALAXIES, AND CLUSTERING}

What is the ``present epoch?''  For statistical analyses of
clustering, a useful definition is the redshift range beyond which
ignorance of, or inability to correct for, the effects of evolution of
galaxies and their clustering becomes the dominant source of
uncertainty. In this section I examine how the apparent power spectrum
is affected by (1) the mapping between redshift and comoving
coordinate distance, (2) galaxy evolution, and (3) clustering
evolution. Careful analysis will allow us to learn about evolution;
the question here is, when do these effects begin to dominate the apparent
clustering?

\begin{figure}
\epsfxsize=12.5cm 
\epsfbox{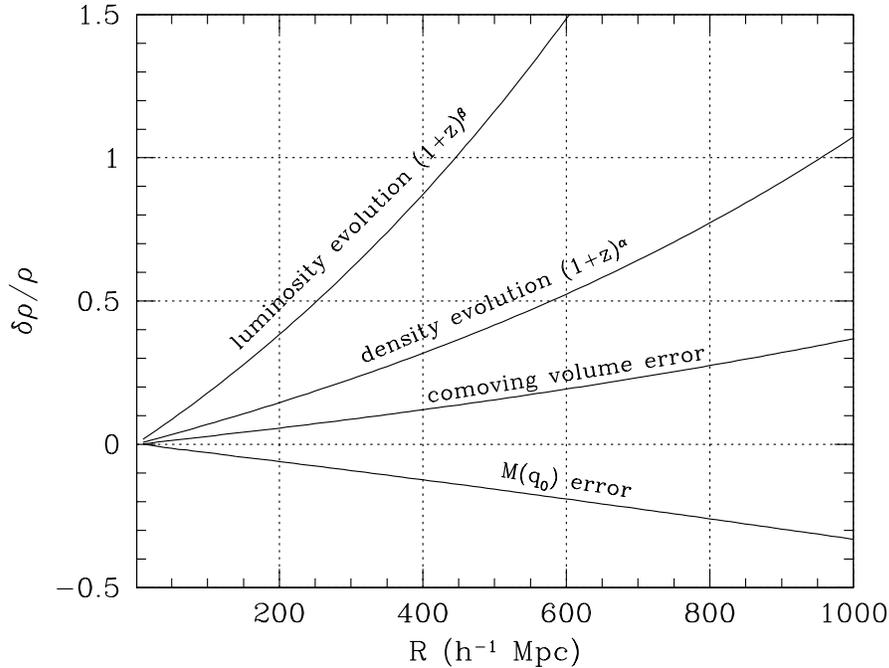}
\vskip-0.5cm
\caption{Density gradients caused by (from top to bottom) galaxy
luminosity evolution with $\beta=2.3$, density evolution with
$\alpha=2$, error in the computed
comoving volume, and error in the computed absolute magnitudes in an
absolute-magnitude limited sample of the SDSS.  The latter two
cosmological effects are shown for the case in which the true
$q_0=0.1$ but we have used $q_0=0.5$ to relate redshifts to comoving
coordinate distances.  }
\end{figure}

\subsection{KNOW THY COSMOLOGY}

The relationship between redshift and comoving coordinate distance 
depends on the deceleration parameter $q_0$,
\begin{equation}
r(z)={c\over H_0 q_0^2(1+z)}\left [q_0z+(1-q_0)\left(1-\sqrt{1+2q_0z}
\right)\right].
\end{equation}
Out to redshifts of a few thousand km~s$^{-1}$, it matters very little
which value of $q_0$ we use.  But at larger distance, this
transformation begins to affect galaxy clustering in several ways.
Using the wrong $q_0$ will cause fluctuations on a fixed
comoving scale to appear as fluctuations on different apparent scales
as a function of redshift. Over the redshift range probed by the SDSS
this effect would cause only minor smearing of features in the spectrum;
varying $q_0$ from $0.1$ to $0.5$ 
changes the length scale by only $4\%$ over the redshift range $z=0$
to $z=0.1$.

It also follows that the comoving volume element depends on $q_0$, as
\begin{equation}
{dV\over dz} = 2\pi \left({c\over H_0}\right)^3 
{\left[q_0z+(1-q_0)(1-\sqrt{1+2q_0z})\right]^2 \over
q_0^4 (1+z)^3\sqrt{1+2q_0z}}.
\end{equation}
If we assume a value of $q_0$ that is too large, the volume element
$dV$ will erroneously decrease with distance, and cause a radial
gradient in the apparent galaxy density. Varying $q_0$ from $0.1$ to $0.5$
raises the comoving galaxy density by $8\%$ at redshift $z=0.1$.

Because construction of volume-limited samples relies on using the
redshift to compute the absolute magnitudes of the galaxies, an error
in $q_0$ translates into an error in the inferred absolute magnitudes,
\begin{equation}
\Delta M = 5\log\left [ r(z,q_0^{true}) / r(z,q_0^{assumed})\right ].
\end{equation}
This error in the absolute magnitudes yields a gradient in apparent galaxy
density, as described by equation 2.
Using too large a value of $q_0$, we would infer that galaxies are fainter
than their true luminosity and exclude some from our sample, thus lowering
the galaxy density at large redshift. At redshift $z=0.1$, varying
$q_0$ from $0.1$ to $0.5$ lowers the density by $7\%$.

Figure 9 shows the density gradients in the SDSS-500 volume-limited
sample that would result from assuming $q_0=0.5$ if the true value is
$q_0=0.1.$   Interestingly, the rise in apparent density that results
from the change in comoving volume element nearly cancels the
absolute-magnitude effect (this cancellation is not universally true
because the absolute-magnitude effect depends on the slope of the
integrated luminosity function at the absolute magnitude limit).

\begin{figure}
\epsfxsize=12.5cm 
\epsfbox{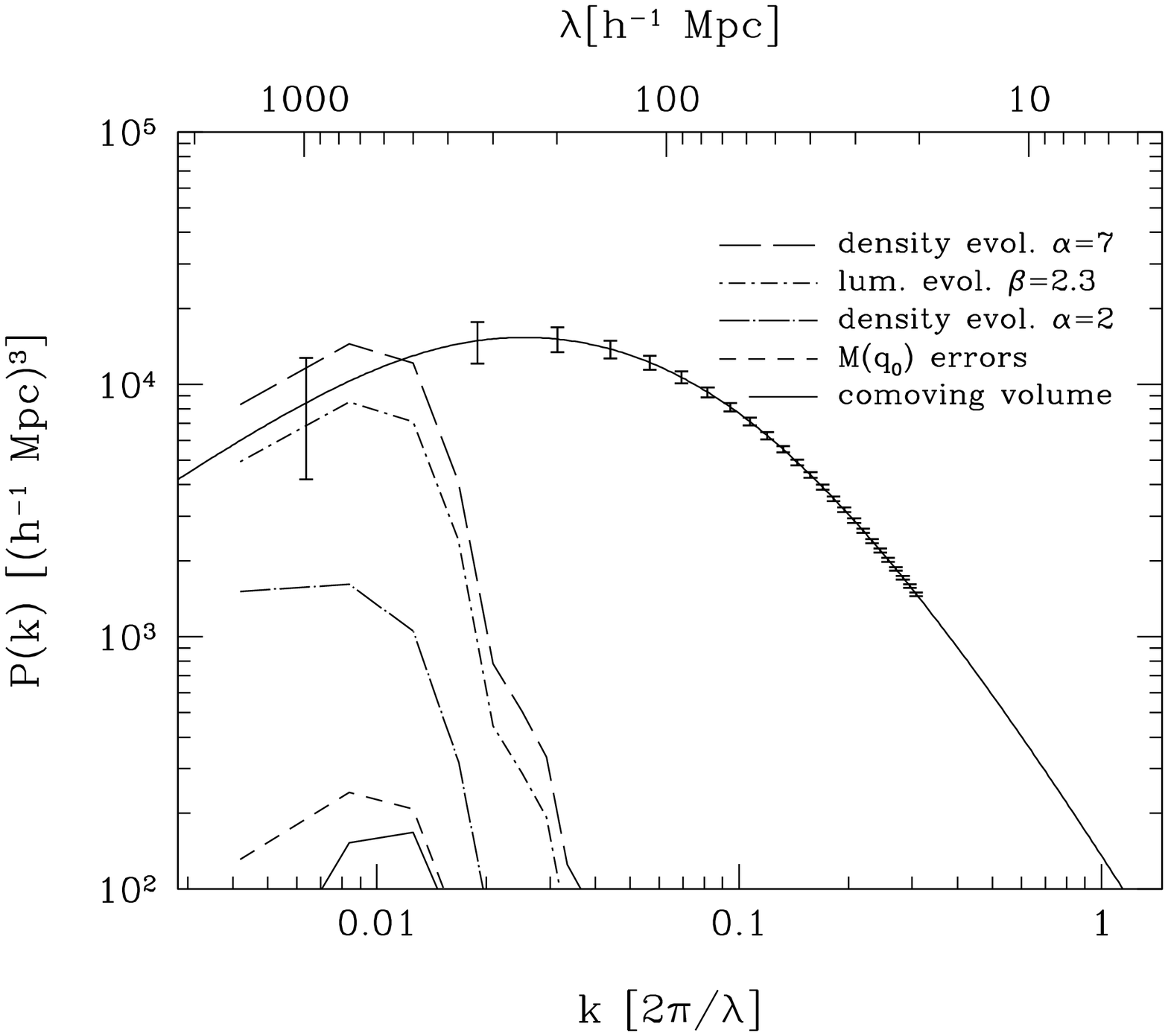}
\vskip-0.5cm
\caption{
False clustering power that would be observed in the SDSS-500
volume-limited sample by (top to bottom) galaxy density evolution with
$\alpha=7$, galaxy luminosity evolution with $\beta=2.3$, density
evolution with $\alpha=2$, and from gradients caused by using the wrong
value of $q_0$.  
In every case, the sharp turnover at small $k$ is
an artifact of our failure to correct for a numerical effect.
}
\end{figure}

Figure 10 illustrates the extra clustering power that results from
these effects, considered separately. Both the comoving volume effect
and the absolute magnitude effect cause negligible power because the
gradient is so shallow.  Even if these gradients did not cancel, they
would not be a cause for worry in samples that extend to $R=500h^{-1}$
Mpc.

\subsection{GALAXY EVOLUTION}

Also shown in Figures 9 and 10 are the density gradients and extra
clustering power predicted for galaxy luminosity or density evolution.
Here we model density evolution as $\bar{n}(z)\propto (1+z)^{\alpha}$
with $\alpha=2$ shown in Figure 9, and both $\alpha=7$ (upper curve)
and $\alpha=2$ (middle curve) in Figure 10. If $\alpha$ is much larger
than $2$, this effect could cause trouble in the SDSS samples. For
example, Saunders \etal~(1990) find evidence for $\alpha=7$ in the
IRAS-selected galaxy luminosity function.  We model luminosity
evolution as $L(z)\propto (1+z)^{\beta}$ with $\beta=2.3$, as
fit by Lilly \etal~ (1995) to their $I$-band luminosity
functions. Saunders \etal~ find $\beta=3$ for IRAS
galaxies. Figure 10 shows that this type of luminosity evolution would
contribute a significant amount of clustering power on the largest
scales probed by the SDSS-500 sample.

A related effect is the difficulty of computing the K-corrections for
galaxies over a large range of redshift. Out to redshift
$z=0.2$, failing to perform the proper K-corrections in $r$' (the
selection band for SDSS spectroscopy) causes an effect on the galaxy
samples that is somewhat smaller than the luminosity evolution effect.

The lesson here is that one cannot treat galaxies as indistinguishable
``points'' for the purpose of power spectrum analysis.  We already
know this from, for example, luminosity bias, whereby very bright
galaxies are more strongly clustered than the average (\eg, Park
\etal~1994).  As we push to larger redshift, we must first characterize
the redshift dependence of the intrinsic properties of galaxies. Only
then can we compute with confidence the clustering statistics of
samples that are either homogeneous in their clustering properties or
that have known dependence of clustering on galaxy species.  Computing
the requisite multivariate luminosity functions and spectral evolution
of galaxies requires multiband photometry and homogeneous samples with
good resolution spectra. The SDSS will provide both of these
desiderata.

\subsection{CLUSTERING EVOLUTION}

Clustering of galaxies fails to be ergodic as we probe deeper into the
universe for yet another reason: the pattern of clustering itself
evolves with time. In linear theory, the growing-mode pertubations
grow with the scale factor $a$ of the universe such that
$P(k,a)\propto D^2(a)$, where $D(a)$ is a function of the
cosmic matter density $\Omega_M$ and cosmological constant
$\Omega_{\Lambda}$.
Relative to the present epoch, when $D(z)\equiv 1$, the growth
factor when the universe had relative scale size
$a \equiv (1+z)^{-1}$ is (Carroll, Press, \& Turner 1992)
\begin{eqnarray}
D(a) & = & {5\over 2}{\Omega_M\over a}\left[1+\Omega_M
\left ({1\over a}-1\right)+\Omega_{\Lambda} (a^2-1)\right]^{1/2} \\
\times & \int^a_0 & \left[1+\Omega_M
\left({1\over a'}-1\right)+\Omega_{\Lambda}(a'^2-1)
\right]^{-3/2} da'. \nonumber
\end{eqnarray}
For $\Omega_M=1$ and
$\Omega_{\Lambda}=0$, this relation is simply $D(z)=(1+z)^{-1}$.

\begin{figure}
\epsfxsize=12.5cm 
\epsfbox{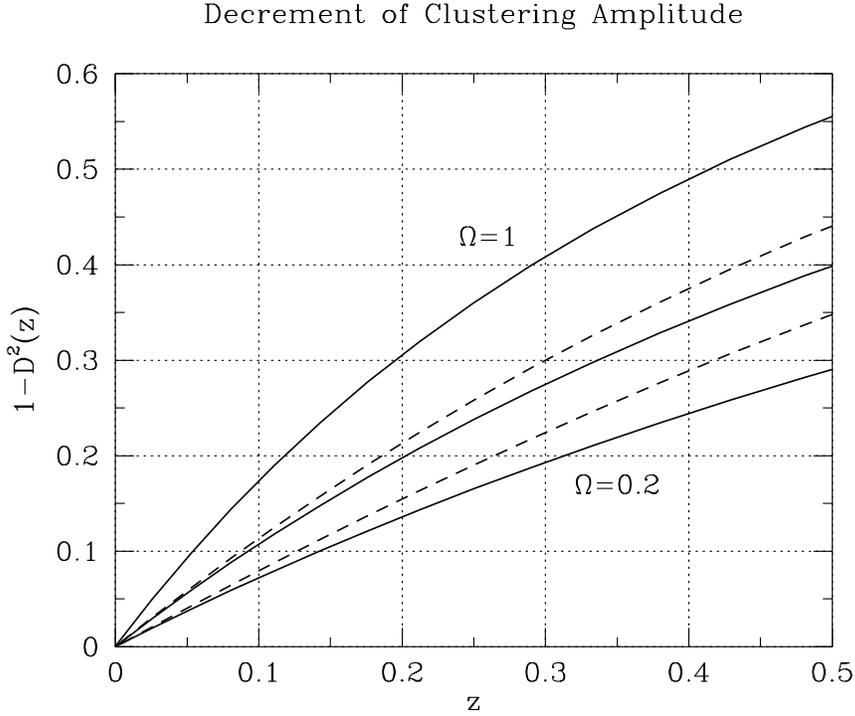}
\vskip-1cm
\caption{Evolution of the clustering amplitude with redshift for
(top to bottom) 
\{$\Omega_M=1$, $\Omega_{\Lambda}=0$\},
\{$\Omega_M=0.4$, $\Omega_{\Lambda}=0.6$\},
\{$\Omega_M=0.4$, $\Omega_{\Lambda}=0$\},
\{$\Omega_M=0.2$, $\Omega_{\Lambda}=0.8$\}, and
\{$\Omega_M=0.2$, $\Omega_{\Lambda}=0$\}.
Plotted here is the difference in clustering amplitude relative to
$z=0$.
}
\end{figure}

Figure 11 plots the apparent decrement of clustering amplitude
relative to $z=0$, $D^2(z=1)-D^2(z)=1-D^2(z)$, for a few choices of
spatially flat and open cosmologies. By redshift $z=0.2$, the
decrement is $30\%$ for $\Omega_M=1$ and $12\%$ for an $\Omega_M=0.2$
open universe. The difference in the decrement for these models is
then larger than the statistical errors in the power spectrum on most
scales for a SDSS sample to this depth.  

In other words, the cosmology-dependence of the growth factor becomes
the dominant source of uncertainty in estimation of the power spectrum
and we reach a fundamental limit on our ability to measure the power
spectrum in an model-independent fashion.  This is not a problem when
comparing with models that specify the cosmology, because one simply
predicts the power spectrum that would be observed over the specified
volume for each model.  But it is not possible to compare local
measures of clustering with observations at larger redshift without
specifying the cosmology. For example, apparent variation in the
clustering amplitude of a galaxy species that are due to galaxy
evolution might be degenerate with variation of the cosmology.
Thus ends the ``present epoch'' for the purpose of statistical large-scale
structure.

On the other hand, given the large divergence of clustering amplitude
for models that might otherwise match at $z=0$, the strong
cosmology-dependence of the growth factor seems like a powerful way to
constrain $\Omega_M$ and $\Omega_{\Lambda}$.  Given all of the caveats
above regarding galaxy evolution, we must find a sample of objects
that (1) has negligible (or, at least, very well understood)
intrinsic evolution out to a redshift of a few tenths (2) are
sufficiently numerous that shot noise does not dominate over the
cosmic variance, and for which we (3) can easily obtain redshifts
(either spectroscopic or photometric).  One must use a single
volume-limited sample, because the decrease of clustering amplitude
with redshift might be masked by the effects of luminosity bias
(brighter galaxies cluster more strongly) in a magnitude-limited
sample. A good candidate for such an analysis will be the BRG sample
of the SDSS, which will extend to $z=0.45$. The brightest, reddest
galaxies evolve the least and have strong spectral features that ease
redshift determination for these objects.  Dividing this redshift
sample into broad bins of redshift and averaging the power estimates
over broad bands should allow determination of the clustering
amplitude to within a few percent at each redshift.

\section{CONCLUSIONS}

The large sample volume and dense sampling of the galaxy distribution
in the SDSS redshift survey will set lower bounds to the uncertainties
that could, in principle, allow detection of small deviations from
smooth spectra and easily differentiate between similar cosmological
models.  Clever use of the anisotropy of clustering would allow
simultaneous estimation of both $\Omega_M$ and the real-space power
spectrum.  Further, we could use the growth rate of clustering to
constrain $\Omega_M$ and $\Omega_{\Lambda}$ in a different
way. However, all of these dreams hinge on our ability to cleanly
separate true galaxy clustering from contamination by extinction and
photometric error, and to differentiate between clustering evolution
and evolution of the galaxies themselves.

Thus, the future of large-scale structure lies in studying evolution,
both of galaxies and of the clustering pattern itself. This evolution
is strongly species-dependent and will require estimation of
multivariate luminosity functions to interpret.  In a few years' time,
the SDSS, together with the 2MASS infrared survey and FIRST radio
surveys, which will also cover (really, have already begun to cover)
the same region of sky, will provide a database of optical, near-IR,
and radio information about millions of objects and will be ideal for
just this sort of investigation.

The author acknowledges the collaborative effort of many of the
participants in the SDSS, in particular Andy Connolly and Alex Szalay,
that contributed to the results presented in this volume.  Support
for this work was provided by NASA through grant HF-01078.01-94A from
the Space Telescope Science Institute, which is operated by AURA,
Inc. under NASA contract NAS5-26555,

\end{document}